\documentclass[aps,prl,twocolumn,groupedaddress,latterpaper]{revtex4}
\usepackage{graphicx}

\begin{document}

\title{A study of random laser modes in disordered photonic crystals}
\author{Alexey Yamilov}
\email{a-yamilov@northwestern.edu}
\author{Hui Cao}
\affiliation{Department of Physics and Astronomy, Northwestern University, Evanston, IL 60208}
\date{\today}

\begin{abstract}

We studied lasing modes in a disordered photonic crystal. The scaling of the lasing threshold with the system size depends on the strength of disorder. For sufficiently large size, the minimum of the lasing threshold occurs at some finite value of disorder strength. The highest random cavity quality factor was comparable to that of an intentionally introduced single defect. At the minimum, the lasing threshold showed a super-exponential decrease with the size of the system. We explain it through a migration of the lasing mode frequencies toward the photonic bandgap center, where the localization length takes the minimum  value. Random lasers with exponentially low thresholds are predicted.

\end{abstract}


\maketitle

A finite open system of scattering particles can be characterized by a set of quasi-stationary (leaky) optical modes. When  gain, sufficient to compensate the leakage in at least one mode, is introduced in such system - random laser is formed \cite{Letokhov,CaoRandomLaserPRL,WiersmaNatureRandomLaser,VardenyPolymerRandomLaser,ModeCompetition,Burin1dRandomLaser,LawandyTheoryRandLaser,GenackDiffRandomLaser,WiersmaLagendijkDiffRandLaser}. Unconventional feedback mechanism in a random laser leads to properties interesting from both fundamental and practical point of view \cite{Letokhov,CaoRandomLaserPRL,WiersmaNatureRandomLaser,VardenyPolymerRandomLaser,ModeCompetition,Burin1dRandomLaser,LawandyTheoryRandLaser,GenackDiffRandomLaser,WiersmaLagendijkDiffRandLaser,JiangSouk,SebbahFDTD,JohnRandLaser,BurinOrder}. While  previous studies concentrated mainly on disordered systems, the resent experiments by Shkunov \textit{et al} \cite{VardenyPhotonicLasing} demonstrated random lasing peaks associated with the bandgap in partially ordered system. In this letter, we obtain the lasing threshold for a random laser with different degree of ordering. Important information about these states can be extracted from the scaling of the average lasing threshold with the system size \cite{BurinOrder}, $L$. For increasing values of the disorder parameter we find 5 scaling regimes: (a) photonic band-edge, $1/L^3$, (b) transitional super-exponential, (c) bandgap-related exponential, (d) diffusive, $1/L^2$, and (e) disorder-induced exponential, due to Anderson localized modes, regimes. In this letter we show that by optimizing the disorderness of the sample one can dramatically reduce the threshold of random laser, to the values comparable to those of photonic bandgap defect lasers \cite{HighQDefectCavity}. This finding can open up a road to practical applications of random lasers.

Finding the lasing threshold of a random laser theoretically is a difficult problem. In disordered photonic crystals one cannot take simplifying assumptions such as independent scattering approximation (low density limit), neglect the finite size of the scatterers, or the confined dimensions of the system. Random laser modes have the highest $Q$'s and therefore weakly contribute to the transport properties of the system. Among other methods \cite{AsatryanOrderToDisorder,NojimaOpenSystem}, finite difference time domain (FDTD) method has been shown to be a convenient tool in studying random laser modes in 1D \cite{JiangSouk} and 2D \cite{SebbahFDTD}. It was recently shown \cite{SebbahFDTD,JiangSouk} that above the threshold, the shape of the lasing modes remain the same as in passive medium. In this work, we used FDTD method to find the modes with the smallest energy decay rate in an open passive 2D random medium system with different degree of order. These modes determine the value of the lasing threshold in an active medium. After an initial pulsed excitation with the frequency $\omega_e$ in vicinity of the photonic bandgap, electric energy of the system followed ${\cal E}\propto Re\; [\exp{2i\omega_m(1+i/2Q_m)}]$ dependence. $\omega_m$ and $Q_m$ were identified as the frequency and quality factor of the random cavity. In what follows, we will refer to $1/Q_m$ as the lasing threshold of the system. Our main results can be summarized as follows: (i) for sufficiently wide bandgaps, the minimum of the lasing threshold develops at some finite value of disorder parameter; (ii) at this ``optimal'' disorderness, the cavity $Q$ is determined by the localization length similar to that of single defect in the ordered sample, leading to a similar random cavity $Q$; (iii) with an increase of the system size the optimal disorder decreases; (iv) near the minimum, lasing threshold scales super-exponentially on sample size, which we ascribe to the migration of mode frequency toward the bandgap center and, therefore the decrease of the ``typical'' localization length.

We consider a 2D photonic crystal made of $N\propto L^2$ cylinders with diameter $d=98 nm$ and refractive index $n_0=2.2$. The cylinders were arranged into hexagonal lattice with nearest-neighbor distance of $a=140 nm$. In the absence of disorder, the infinite system with these parameters has full bandgap in $[361nm,426nm]$ range for TM modes (electric field along the cylinder axis). The disorder in the system was introduced in 2 ways: by uniformly randomizing the refractive index of different cylinders in the range $[n_0-w_n(n_0-1),n_0+w_n(n_0-1)]$ and diameter - $[d(1-w_d),d(1+w_d)]$. Care was taken in order to avoid the uncontrollable disorder due to discretization of the grid. Disorder in the system was characterized with parameter $\delta \varepsilon =<\int (\varepsilon(\mathbf{r})- \varepsilon_0(\mathbf{r}))^2d\mathbf{r}>^{1/2}/(\int \varepsilon_0^2(\mathbf{r})d\mathbf{r})^{1/2}$, where $\varepsilon_0(\mathbf{r})$ and $\varepsilon(\mathbf{r})$ is the dielectric constant distribution in ordered and disordered samples respectively. $<...>$ stands for the average over different disorder configurations. In the present work we studied the systems with $11$ different disorder strengths: $1$ to $10$ had $w_n\in[0.1,1.0]$ with the increment $0.1$ and $w_d\equiv 0$, the $11$th had $w_n=1.0$ and  $w_d=0.43$. This led to variations of dielectric constant, $\delta\varepsilon$, from weak - $0.08$ to strong - $0.95$ disorder. Later we will discuss the effects of this particular choice of these types of the disorder. 
 We used standard FDTD method to study the evolution of TM electromagnetic field in the described system system with $\lambda/20$ resolution grid. In the studied range of system sizes, $N\in[75,600]$, the required grid sizes varied from $200\times 200$ to $500\times 500$. To mimic an open system, a buffer layer of air of $150 nm$ thickness was kept around the sample, followed by uniaxial perfectly matched layer \cite{taflove}. To excite the system we initially launched a short $\sim 10fs$ pulse at every grid point. The frequency of the pulse was at the center of the bandgap, $391nm$, of the ordered structure. In the frequency domain the full width at half maximum of the seed pulse was of the order of the bandgap width. Thus the seed pulse excited all states within the stop band and near the band edges. After initial excitation pulse, the system was left on its own. Field distribution in the system and total electric energy ${\cal E}=1/2\int \varepsilon(\mathbf{r})\mathbf{E}^2(\mathbf{r})d\mathbf{r}$ was monitored.

Right after the initial pulse, the competition between the modes \cite{ModeCompetition,BurinDipoleLaser1} with the different lifetimes led to the complicated evolution of ${\cal E}$. However, after a sufficient time only the mode with the longest lifetime (maximum quality factor) survived. ${\cal E}$ followed a mono-exponential dependence $Re[\exp{2i\omega_m(1+i/2Q_m)}]$, from which we extracted the maximum quality factor $Q_m$ and frequency $\omega_m$ for this particular realization of the disorder. At the same time the spatial pattern, $\mathbf{E}(\mathbf{r})$, stabilized and the mode profile could be seen. 
Generally, the time needed to reach the mono-exponential regime varied from about $0.5ps$ for the smallest system to $10ps$ for the largest. Finally, the $1/Q_m$, representing lasing threshold, was averaged over $1000$ ($N=75$), or $100$ ($N=137,188,261,368,449,608$) disorder realizations.

\begin{figure}
\includegraphics{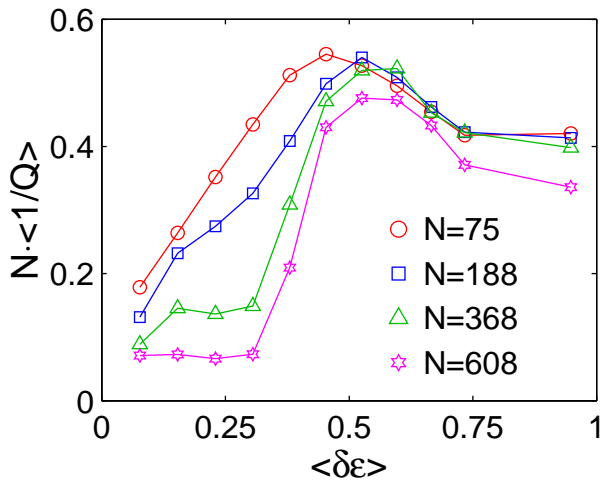}
\caption{\label{thresh} Average inverse random cavity quality factor normalized by total number of scatterers $N\propto L^2$, as a function of disorderness. }
\end{figure}

\begin{figure}
\includegraphics{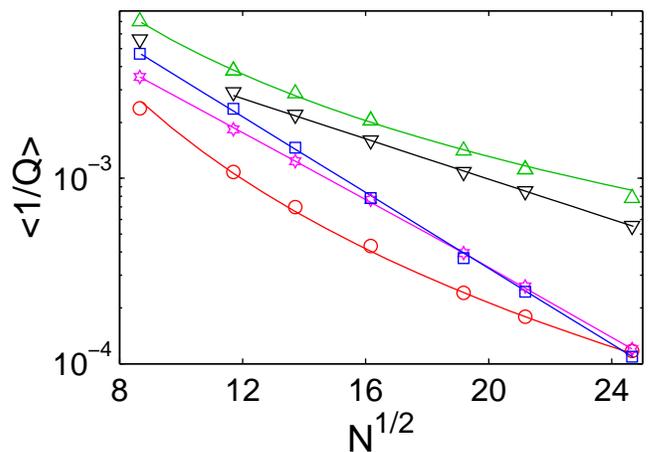}
\caption{\label{size} Scaling of the average inverse quality factor with the lateral size of the system $L\propto N^{1/2}$ in 5 regimes:  $L^{-3}$ band-edge type - circles; super-exponential  - stars (line is used to connect points); bandgap exponential - squares; $L^{-2}$ diffusion type - triangles; and disorder exponential - inverted triangles. These correspond to  
$w_n =0.1\; (\delta\varepsilon\simeq 0.08)$,  
$w_n =0.2\; (\delta\varepsilon\simeq 0.15)$,  
$w_n =0.3\; (\delta\varepsilon\simeq 0.23)$,  
$w_n =0.7\; (\delta\varepsilon\simeq 0.53)$, and  
$w_n =1.0\; \&\;w_d=0.43\; (\delta\varepsilon\simeq 0.95)$ values of disorder parameter respectively.}
\end{figure}

Fig. \ref{thresh} shows the dependence of the average lasing threshold normalized by $N$ as a function of the disorderness, different curves correspond to different system sizes. This particular normalization makes it easy to see the deviation from diffusion predicted \cite{Letokhov} dependence $1/L^2\propto N$. One can see that significantly different scaling at different $\delta\varepsilon$ leads to the appearance of a minimum at the finite disorder strength. Fig. \ref{size} plots $L$ dependence of the threshold for the fixed disorder strengths. 

\begin{figure}
\includegraphics{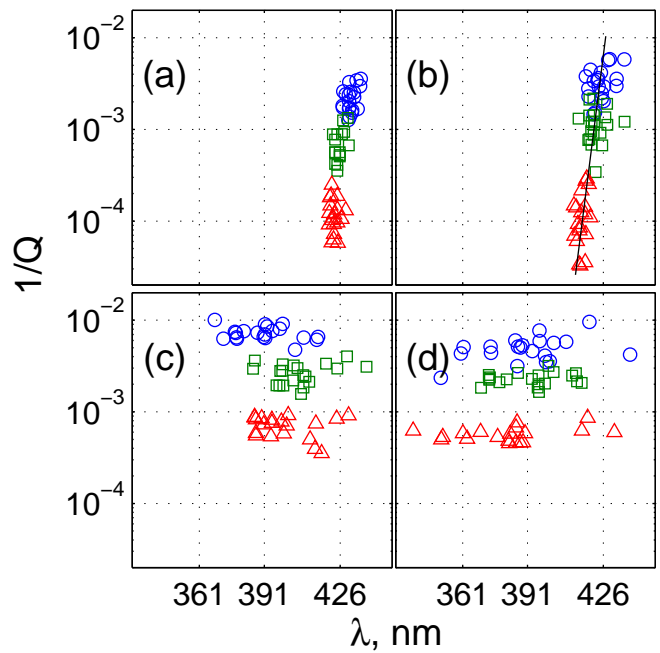}
\caption{\label{dist} $1/Q$ vs cavity mode frequency for 20 realizations of disorder. Circles, squares, and triangles correspond to $N$ equal to 75, 188, and 608 respectively. Four graphs correspond to different disorder parameters:  $w_n =0.1\; (\delta\varepsilon\simeq 0.08)$ - (a), $w_n =0.2\; (\delta\varepsilon\simeq 0.15)$ - (b), $w_n =0.6\; (\delta\varepsilon\simeq 0.45)$ - (c), $w_n =1.0\; \&\;w_d=0.43\; (\delta \varepsilon\simeq 0.95)$  - (d).}
\end{figure}

The understanding of this behavior comes from the position of the mode frequencies - Fig. \ref{dist}. For small $\delta \varepsilon$ the frequencies are concentrated at lower (long-wavelength) band-edge, and they (as well as $Q_m$) are independent on the position $\omega_e$ of the excitation pulse. The reason for that is the way the disorder was introduced into the system. The long-wavelength modes are mostly concentrated in the high dielectric medium, which is being disordered by the index fluctuations. At $w_n=0.1$ the mode frequencies fell  in the immediate vicinity of the band-edge, Fig. \ref{dist}a. Lasing from the band-edge modes is well studied in the case of the ordered structures \cite{DowlingBandEdgeLaser,SakodaBandEdgeScaling,VictorAziBandEdgeLaser,SusaBandEdge,LeeBandEdgeExp,BurinOrder}, with $1/L^3$ dependence of the threshold \cite{
BurinOrder} giving a good fit with our results. At the increased disorder, $w_n=0.2$, we observe the decrease of the threshold that is noticeably faster than exponential, expected for localized modes \cite{Burin1dRandomLaser}. This unexpected behavior becomes clear from Fig. \ref{dist}b. The value of the threshold can be estimated as $1/Q\propto \exp(- L/2\xi(\omega_m))$, where $\xi (\omega _m)$ is some ``typical'' value of the localization length at the frequency $\omega_m$. From Fig. \ref{dist}b one can see that with increased size, the frequency of the modes advance toward the band center, where $\xi$ is the shortest, magnifying the exponential dependence of the threshold. The explanation for this frequency migration with the increase of the size comes from the fact that in the finite system the modes deep into the bandgap are not accessible due to low density of states there. The Urbach-like behavior can be expected \cite{AsatryanOrderToDisorder,PingShengTail}. The support for this, indirectly comes from Fig. \ref{dist}b, where $1/Q$ follows rather well the exponential dependence (solid line) on the frequency shift from the band-edge of the ordered system. However, because of the special nature of the modes no direct conclusions can be made. Assuming Urbach-like dependence of the density of states, the advancement of the mode frequency can be estimated from the condition that total number of states, $N$, times the probability of having a state located $\Delta \omega(N)$ away from the band-edge, $\exp (-\alpha (\delta\varepsilon)\cdot \Delta \omega(N))$, is still of order of one. Here, $\alpha (\delta\varepsilon)$ is exponential slope in the density of states, that should decrease with the increase of the disorder $\delta\varepsilon$. For small disorders $\alpha^{-1}(\delta\varepsilon) \ll \Delta E_{PBG}$, where $\Delta E_{PBG}$ is the width of the photonic gap. Therefore, for weak disorder (or small system size) the bandedge-type modes have the highest $Q$. The crossover to the super-exponential dependence of the threshold occurs when the cavity $Q$ associated with the localized states with the smallest localization length $\xi(\Delta \omega(N))$ available for this size $N$ exceeds that of the bandedge-type mode: $\exp{[-N^{1/2}a/2\xi(\Delta \omega(N))]}\sim 1/N^{3/2}$. Stronger size dependence in the super-exponential regime means that even $N^{-3/2}$ band-edge type dependence observed at smaller disorder would eventually switch to the super-exponential one. However, the latter can be expected to saturate at larger sizes \textit{or} disorders when the mode frequency reaches the bandgap center: $N\cdot \exp (-\alpha (\delta\varepsilon)\cdot\Delta E_{PBG}/2 )\sim 1$, where the localization length is the smallest. Therefore, we expect the limiting scaling of the lasing threshold to be exponential $\exp{[-N^{1/2}a/2\xi(\Delta E_{PBG}/2)]}$. Three important conclusions (i-iii) listed in the introduction immediately follow. (iii) is justified because $\xi(\Delta E_{PBG}/2)$ is the smallest in the ordered sample. The saturated exponential dependence can be seen in our example for $w_n=0.3$ - squares in Fig. \ref{size}.

\begin{figure}
\includegraphics{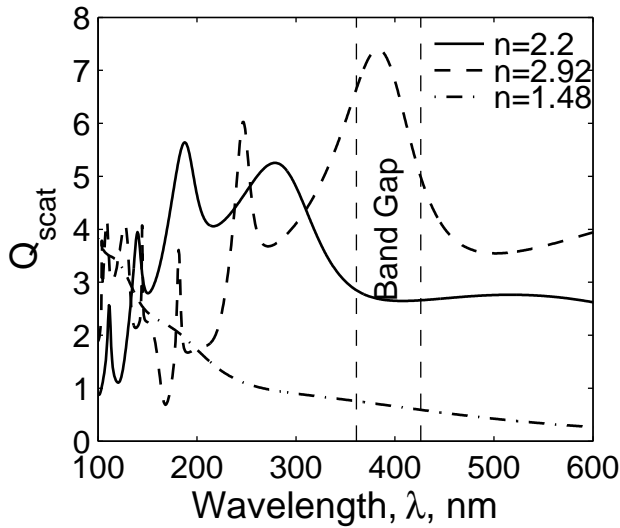}
\caption{\label{scat} Mie scattering efficiency vs wavelength for cylinders of diameter $98 nm$. Refractive index $n=2.2$ - solid line. For $w_n =0.6\; (\delta\varepsilon\simeq 0.45)$, the refractive index fluctuates from $n=1.48$ (dash-dotted line) to $n=2.92$ (dashed line) }
\end{figure}

In our system the sharp rise of the threshold at $\delta \varepsilon \simeq 0.4$ on Fig. \ref{thresh} is attributed to the removal of the bandgap. This could be seen as the loss of the hexagonal symmetry of the observed modes as well a sensitivity of the modes to the excitation pulse position $\omega_e$. In this regime, the frequency of the modes is not associated with the photonic gap, which does not exist anymore. However, in order to make a direct comparison with the ordered case, we kept the excitation pulse the same as before. The exact cause of the disappearance of the bandgap is being debated in the literature \cite{BandGapDestruction,AsatryanOrderToDisorder,PingShengTail}, and is not the subject of this study. In our particular case we found
a simple explanation of the threshold behavior in the way the disorder was introduced. Indeed, the fluctuating index of refraction leads to the fluctuation of the frequency of the Mie resonances of the particle. For box distribution, there exists a value of $w_n=0.6$ when the defect Mie resonance falls into the gap, Fig. \ref{scat}. This value matches the value of disorder parameter, when the sharp increase of the lasing threshold is observed on Fig. \ref{thresh}. Moreover, Fig. \ref{dist}c shows that at this crossover disorder, the modes avoid the region of high scattering, which is the consequence of the sharp boundary in the distribution of $n$. It also indicates the presence of the residual bandgap, where the lasing frequencies are concentrated.

At $w_n\geq 0.7$ the photonic bandgap ceased to exist. Triangles on Fig. \ref{size} correspond to $w_n=0.7$, and were successfully fitted with diffusion \cite{Letokhov,LawandyTheoryRandLaser,GenackDiffRandomLaser,WiersmaLagendijkDiffRandLaser,JohnRandLaser,BurinOrder} scaling dependence $1/L^2$. Deviations from this dependence can be seen from Figs. \ref{thresh},\ref{size} at the largest sizes studied, where $L>\xi_{Anderson}$ the states become localized again by virtue of Anderson localization \cite{Burin1dRandomLaser,BurinOrder}. The exponential dependence becomes especially pronounced at the largest disorder studied (see inverted triangles on Fig. \ref{size}), where the transition from $1/L^2$ to exponential dependence comes at small system sizes. We want to point out that even at such strong disorder, the obtained modes had a collective nature, rather than the single particle's (high-order resonances), which were concentrated at higher frequencies. Comparing the localization length of these states to that of bandgap nature we see the factor of two difference, which makes the latter preferable - Fig. \ref{thresh}. This is in accordance with experiments of Ref. \cite{VardenyPhotonicLasing}.

In conclusion, we studied the modes responsible for the lasing threshold of a random laser based on disordered photonic crystal with a complete bandgap. At different disorder strengths, we identified 5 scaling regimes of the lasing threshold. Difference in scaling rates led to the development of the minimum at ``optimal'' disorder, where the threshold takes minimum value. We, for the first time, showed  the regime of super-exponential scaling, which was attributed to the migration of the lasing frequencies to the center of the bandgap. We also predict that the value of the ``optimal'' disorder should decrease with increase of the volume of the system. These results can be generalized to the case of the 3D disordered photonic crystals.

\begin{acknowledgments}
A. L. Burin is acknowledged for fruitful discussions. This work is supported by the National Science Foundation under Grant No DMR-0093949. HC acknowledges the support from the David and Lucile Packard Foundation and Alfred P. Sloan Foundation.
\end{acknowledgments}

\bibliography{disphc}

\begin{thebibliography}{26}
\expandafter\ifx\csname natexlab\endcsname\relax\def\natexlab#1{#1}\fi
\expandafter\ifx\csname bibnamefont\endcsname\relax
  \def\bibnamefont#1{#1}\fi
\expandafter\ifx\csname bibfnamefont\endcsname\relax
  \def\bibfnamefont#1{#1}\fi
\expandafter\ifx\csname citenamefont\endcsname\relax
  \def\citenamefont#1{#1}\fi
\expandafter\ifx\csname url\endcsname\relax
  \def\url#1{\texttt{#1}}\fi
\expandafter\ifx\csname urlprefix\endcsname\relax\def\urlprefix{URL }\fi
\providecommand{\bibinfo}[2]{#2}
\providecommand{\eprint}[2][]{\url{#2}}

\bibitem[{\citenamefont{Letokhov}(1968)}]{Letokhov}
\bibinfo{author}{\bibfnamefont{V.~S.} \bibnamefont{Letokhov}},
  \bibinfo{journal}{Sov. Phys. JETP} \textbf{\bibinfo{volume}{26}},
  \bibinfo{pages}{835} (\bibinfo{year}{1968}).

\bibitem[{\citenamefont{Cao et~al.}(1999)\citenamefont{Cao, Zhao, Ho, Seelig,
  Wang, and Chang}}]{CaoRandomLaserPRL}
\bibinfo{author}{\bibfnamefont{H.}~\bibnamefont{Cao}},
  \bibinfo{author}{\bibfnamefont{Y.}~\bibnamefont{Zhao}},
  \bibinfo{author}{\bibfnamefont{S.~T.} \bibnamefont{Ho}},
  \bibinfo{author}{\bibfnamefont{E.~W.} \bibnamefont{Seelig}},
  \bibinfo{author}{\bibfnamefont{Q.~H.} \bibnamefont{Wang}}, \bibnamefont{and}
  \bibinfo{author}{\bibfnamefont{R.~P.~H.} \bibnamefont{Chang}},
  \bibinfo{journal}{Phys. Rev. Lett.} \textbf{\bibinfo{volume}{82}},
  \bibinfo{pages}{2278} (\bibinfo{year}{1999}).

\bibitem[{\citenamefont{Wiersma}(2000)}]{WiersmaNatureRandomLaser}
\bibinfo{author}{\bibfnamefont{D.~S.} \bibnamefont{Wiersma}},
  \bibinfo{journal}{Nature} \textbf{\bibinfo{volume}{406}},
  \bibinfo{pages}{132} (\bibinfo{year}{2000}).

\bibitem[{\citenamefont{Frolov et~al.}(1999)\citenamefont{Frolov, Vardeny, and
  Yoshino}}]{VardenyPolymerRandomLaser}
\bibinfo{author}{\bibfnamefont{S.~V.} \bibnamefont{Frolov}},
  \bibinfo{author}{\bibfnamefont{Z.~V.} \bibnamefont{Vardeny}},
  \bibnamefont{and} \bibinfo{author}{\bibfnamefont{K.}~\bibnamefont{Yoshino}},
  \bibinfo{journal}{Phys. Rev. B} \textbf{\bibinfo{volume}{57}},
  \bibinfo{pages}{9141} (\bibinfo{year}{1999}).

\bibitem[{\citenamefont{Misirpashaev and Beenakker}(1998)}]{ModeCompetition}
\bibinfo{author}{\bibfnamefont{T.~S.} \bibnamefont{Misirpashaev}}
  \bibnamefont{and} \bibinfo{author}{\bibfnamefont{C.~W.~J.}
  \bibnamefont{Beenakker}}, \bibinfo{journal}{Phys. Rev. A}
  \textbf{\bibinfo{volume}{57}}, \bibinfo{pages}{2041} (\bibinfo{year}{1998}).

\bibitem[{\citenamefont{Burin et~al.}(2002{\natexlab{a}})\citenamefont{Burin,
  Ratner, Cao, and Chang}}]{Burin1dRandomLaser}
\bibinfo{author}{\bibfnamefont{A.~L.} \bibnamefont{Burin}},
  \bibinfo{author}{\bibfnamefont{M.~A.} \bibnamefont{Ratner}},
  \bibinfo{author}{\bibfnamefont{H.}~\bibnamefont{Cao}}, \bibnamefont{and}
  \bibinfo{author}{\bibfnamefont{S.~H.} \bibnamefont{Chang}},
  \bibinfo{journal}{Phys. Rev. Lett.} \textbf{\bibinfo{volume}{88}},
  \bibinfo{pages}{093904} (\bibinfo{year}{2002}{\natexlab{a}}).

\bibitem[{\citenamefont{Balachandran et~al.}(1997)\citenamefont{Balachandran,
  Lawandy, and Moon}}]{LawandyTheoryRandLaser}
\bibinfo{author}{\bibfnamefont{R.~M.} \bibnamefont{Balachandran}},
  \bibinfo{author}{\bibfnamefont{N.~M.} \bibnamefont{Lawandy}},
  \bibnamefont{and} \bibinfo{author}{\bibfnamefont{J.~A.} \bibnamefont{Moon}},
  \bibinfo{journal}{Opt. Lett.} \textbf{\bibinfo{volume}{22}},
  \bibinfo{pages}{319} (\bibinfo{year}{1997}).

\bibitem[{\citenamefont{Siddique et~al.}(1996)\citenamefont{Siddique, Alfano,
  Berger, Kempe, and Genack}}]{GenackDiffRandomLaser}
\bibinfo{author}{\bibfnamefont{M.}~\bibnamefont{Siddique}},
  \bibinfo{author}{\bibfnamefont{R.~R.} \bibnamefont{Alfano}},
  \bibinfo{author}{\bibfnamefont{G.~A.} \bibnamefont{Berger}},
  \bibinfo{author}{\bibfnamefont{M.}~\bibnamefont{Kempe}}, \bibnamefont{and}
  \bibinfo{author}{\bibfnamefont{A.~Z.} \bibnamefont{Genack}},
  \bibinfo{journal}{Opt. Lett.} \textbf{\bibinfo{volume}{21}},
  \bibinfo{pages}{450} (\bibinfo{year}{1996}).

\bibitem[{\citenamefont{Wiersma and
  Lagendijk}(1996)}]{WiersmaLagendijkDiffRandLaser}
\bibinfo{author}{\bibfnamefont{D.~S.} \bibnamefont{Wiersma}} \bibnamefont{and}
  \bibinfo{author}{\bibfnamefont{A.}~\bibnamefont{Lagendijk}},
  \bibinfo{journal}{Phys. Rev. E} \textbf{\bibinfo{volume}{54}},
  \bibinfo{pages}{4256} (\bibinfo{year}{1996}).

\bibitem[{\citenamefont{Jiang and Soukoulis}(2000)}]{JiangSouk}
\bibinfo{author}{\bibfnamefont{X.~Y.} \bibnamefont{Jiang}} \bibnamefont{and}
  \bibinfo{author}{\bibfnamefont{C.~M.} \bibnamefont{Soukoulis}},
  \bibinfo{journal}{Phys. Rev. Lett.} \textbf{\bibinfo{volume}{85}},
  \bibinfo{pages}{70} (\bibinfo{year}{2000}).

\bibitem[{\citenamefont{Vanneste and Sebbah}(2001)}]{SebbahFDTD}
\bibinfo{author}{\bibfnamefont{C.}~\bibnamefont{Vanneste}} \bibnamefont{and}
  \bibinfo{author}{\bibfnamefont{P.}~\bibnamefont{Sebbah}},
  \bibinfo{journal}{Phys. Rev. Lett.} \textbf{\bibinfo{volume}{87}},
  \bibinfo{pages}{183903} (\bibinfo{year}{2001}).

\bibitem[{\citenamefont{Burin et~al.}(2002{\natexlab{b}})\citenamefont{Burin,
  Ratner, Cao, Schatz, and Chang}}]{BurinOrder}
\bibinfo{author}{\bibfnamefont{A.~L.} \bibnamefont{Burin}},
  \bibinfo{author}{\bibfnamefont{M.~A.} \bibnamefont{Ratner}},
  \bibinfo{author}{\bibfnamefont{H.}~\bibnamefont{Cao}},
  \bibinfo{author}{\bibfnamefont{G.~C.} \bibnamefont{Schatz}},
  \bibnamefont{and} \bibinfo{author}{\bibfnamefont{R.~P.~H.}
  \bibnamefont{Chang}}, \bibinfo{journal}{Phys. Rev. Lett.}
  (\bibinfo{year}{2002}{\natexlab{b}}), \bibinfo{note}{submitted}.

\bibitem[{\citenamefont{John and Pang}(1996)}]{JohnRandLaser}
\bibinfo{author}{\bibfnamefont{S.}~\bibnamefont{John}} \bibnamefont{and}
  \bibinfo{author}{\bibfnamefont{G.}~\bibnamefont{Pang}},
  \bibinfo{journal}{Phys. Rev. A} \textbf{\bibinfo{volume}{54}},
  \bibinfo{pages}{3642} (\bibinfo{year}{1996}).

\bibitem[{Var()}]{VardenyPhotonicLasing}
\bibinfo{note}{M. N. Shkunov, M. C. DeLong, M. E. Raikh, Z. V. Vardeny, A. A.
  Zakhidov, and R. P. Baughman, Synthetic Metals \textbf{116}, 485 (2001); M.
  N. Shkunov, Z. V. Vardeny, M. C. DeLong, R. C. Polson, A. A. Zakhidov, and R.
  P. Baughman, Adv. Funct. Mater. \textbf{12}, 21 (2002)}.

\bibitem[{Hig()}]{HighQDefectCavity}
\bibinfo{note}{O. J. Painter, R. K. Lee, A. Scherer, A. Yariv, J. D. O`Brien,
  P. D. Dapkus, and I. Kim, Science \textbf{284}, 1819 (1999); J.-K. Hwang,
  H.-Y. Ryu, D.-S. Song, I.-Y. Han, H.-W. Song, H.-K. Park, Y.-H. Lee, and
  D.-H. Jang, Appl. Phys. Lett. \textbf{76}, 2982 (2000); N. Susa, IEEE J.
  Quant. Electr. \textbf{37}, 1420 (2001); P. R. Villeneuve, S. Fan, and J. D.
  Joannopoulous, Phys. Rev. B \textbf{54}, 7837 (1996); K. Sakoda, J. Appl.
  Phys. \textbf{84}, 1210 (1998); V. Kuzmiak, and A. A. Maradudin, Phys. Rev. B
  \textbf{57}, 15242 (1998); P. Pottier, C. Seassal, X. Letartre, J. L.
  Leclercq, P. Victorovich, D. Cassagne, and C. Jouanin, J. Light-wave Technol.
  \textbf{17}, 2058 (1999); M. Qui, and S. He, Phys. Rev. B \textbf{61}, 12871
  (2000).}

\bibitem[{\citenamefont{McPhedran et~al.}(1999)\citenamefont{McPhedran, Botten,
  Asatryan, N.~A.~Nicorovici, and Robinson}}]{AsatryanOrderToDisorder}
\bibinfo{author}{\bibfnamefont{R.~C.} \bibnamefont{McPhedran}},
  \bibinfo{author}{\bibfnamefont{L.~C.} \bibnamefont{Botten}},
  \bibinfo{author}{\bibfnamefont{A.~A.} \bibnamefont{Asatryan}},
  \bibinfo{author}{\bibfnamefont{C.~M. d.~S.} \bibnamefont{N.~A.~Nicorovici}},
  \bibnamefont{and} \bibinfo{author}{\bibfnamefont{P.~A.}
  \bibnamefont{Robinson}}, \bibinfo{journal}{Aust. J. Phys.}
  \textbf{\bibinfo{volume}{52}}, \bibinfo{pages}{791} (\bibinfo{year}{1999}).

\bibitem[{\citenamefont{Nojima}(2001)}]{NojimaOpenSystem}
\bibinfo{author}{\bibfnamefont{S.}~\bibnamefont{Nojima}},
  \bibinfo{journal}{Appl. Phys. Lett.} \textbf{\bibinfo{volume}{79}},
  \bibinfo{pages}{1959} (\bibinfo{year}{2001}).

\bibitem[{\citenamefont{Taflove and Hagness}(2000)}]{taflove}
\bibinfo{author}{\bibfnamefont{A.}~\bibnamefont{Taflove}} \bibnamefont{and}
  \bibinfo{author}{\bibfnamefont{S.~C.} \bibnamefont{Hagness}},
  \emph{\bibinfo{title}{Computational Electrodynamics}}
  (\bibinfo{publisher}{Artech House}, \bibinfo{address}{Boston},
  \bibinfo{year}{2000}), \bibinfo{edition}{2nd} ed.

\bibitem[{\citenamefont{Burin et~al.}(2001)\citenamefont{Burin, Ratner, Cao,
  and Chang}}]{BurinDipoleLaser1}
\bibinfo{author}{\bibfnamefont{A.~L.} \bibnamefont{Burin}},
  \bibinfo{author}{\bibfnamefont{M.~A.} \bibnamefont{Ratner}},
  \bibinfo{author}{\bibfnamefont{H.}~\bibnamefont{Cao}}, \bibnamefont{and}
  \bibinfo{author}{\bibfnamefont{R.~P.~H.} \bibnamefont{Chang}},
  \bibinfo{journal}{Phys. Rev. Lett.} \textbf{\bibinfo{volume}{87}},
  \bibinfo{pages}{215503} (\bibinfo{year}{2001}).

\bibitem[{\citenamefont{Dowling et~al.}(1994)\citenamefont{Dowling, Scalora,
  Bloemer, and Bowden}}]{DowlingBandEdgeLaser}
\bibinfo{author}{\bibfnamefont{J.~P.} \bibnamefont{Dowling}},
  \bibinfo{author}{\bibfnamefont{M.}~\bibnamefont{Scalora}},
  \bibinfo{author}{\bibfnamefont{M.~J.} \bibnamefont{Bloemer}},
  \bibnamefont{and} \bibinfo{author}{\bibfnamefont{C.~M.}
  \bibnamefont{Bowden}}, \bibinfo{journal}{J. Appl. Phys.}
  \textbf{\bibinfo{volume}{75}}, \bibinfo{pages}{1896} (\bibinfo{year}{1994}).

\bibitem[{\citenamefont{Kopp et~al.}(1998)\citenamefont{Kopp, Fan, Vithana, and
  Genack}}]{VictorAziBandEdgeLaser}
\bibinfo{author}{\bibfnamefont{V.~I.} \bibnamefont{Kopp}},
  \bibinfo{author}{\bibfnamefont{B.}~\bibnamefont{Fan}},
  \bibinfo{author}{\bibfnamefont{H.~K.~M.} \bibnamefont{Vithana}},
  \bibnamefont{and} \bibinfo{author}{\bibfnamefont{A.~Z.}
  \bibnamefont{Genack}}, \bibinfo{journal}{Optics Lett.}
  \textbf{\bibinfo{volume}{23}}, \bibinfo{pages}{1707} (\bibinfo{year}{1998}).

\bibitem[{\citenamefont{Susa}(2001)}]{SusaBandEdge}
\bibinfo{author}{\bibfnamefont{N.}~\bibnamefont{Susa}}, \bibinfo{journal}{Jpn.
  J. Appl. Phys.} \textbf{\bibinfo{volume}{40}}, \bibinfo{pages}{142}
  (\bibinfo{year}{2001}).

\bibitem[{\citenamefont{Ryu et~al.}(2002)\citenamefont{Ryu, Kwon, Lee, Lee, and
  Kim}}]{LeeBandEdgeExp}
\bibinfo{author}{\bibfnamefont{H.-Y.} \bibnamefont{Ryu}},
  \bibinfo{author}{\bibfnamefont{S.-H.} \bibnamefont{Kwon}},
  \bibinfo{author}{\bibfnamefont{Y.-J.} \bibnamefont{Lee}},
  \bibinfo{author}{\bibfnamefont{Y.-H.} \bibnamefont{Lee}}, \bibnamefont{and}
  \bibinfo{author}{\bibfnamefont{J.-S.} \bibnamefont{Kim}},
  \bibinfo{journal}{Appl. Phys. Lett.} \textbf{\bibinfo{volume}{80}},
  \bibinfo{pages}{3476} (\bibinfo{year}{2002}).

\bibitem[{\citenamefont{Sakoda et~al.}(1999)\citenamefont{Sakoda, Ohtaka, and
  Ueta}}]{SakodaBandEdgeScaling}
\bibinfo{author}{\bibfnamefont{K.}~\bibnamefont{Sakoda}},
  \bibinfo{author}{\bibfnamefont{K.}~\bibnamefont{Ohtaka}}, \bibnamefont{and}
  \bibinfo{author}{\bibfnamefont{T.}~\bibnamefont{Ueta}},
  \bibinfo{journal}{Optics Express} \textbf{\bibinfo{volume}{4}},
  \bibinfo{pages}{481} (\bibinfo{year}{1999}).

\bibitem[{\citenamefont{Frigerio et~al.}(1993)\citenamefont{Frigerio, Rivory,
  and Sheng}}]{PingShengTail}
\bibinfo{author}{\bibfnamefont{J.~M.} \bibnamefont{Frigerio}},
  \bibinfo{author}{\bibfnamefont{J.}~\bibnamefont{Rivory}}, \bibnamefont{and}
  \bibinfo{author}{\bibfnamefont{P.}~\bibnamefont{Sheng}},
  \bibinfo{journal}{Opt. Comm.} \textbf{\bibinfo{volume}{98}},
  \bibinfo{pages}{231} (\bibinfo{year}{1993}).

\bibitem[{Ban()}]{BandGapDestruction}
\bibinfo{note}{S. Fan, and P. R. Villeneuve, and J. D. Joannopoulos, J. Appl.
  Phys. \textbf{78}, 1415 (1995); A. A. Asatryan, and P. A. Robinson, and L. C.
  Botten, R. C. McPhedran, and N. A. Nicorovici, and C. Martijin de Sterke,
  Phys. Rev. E \textbf{62}, 5711 (2000); M. M. Sigalas, and C. M. Soukoulis,
  and C. T. Chan, and R. Biswas, and K. M. Ho, Phys. Rev. B \textbf{59}, 12767
  (1999); Z. Y. Li, and Z. Q. Zhang, Phys. Rev. B \textbf{62}, 1516 (2000)}.

\end{thebibliography}

\end{document}